\documentclass[journal, 10pt]{IEEEtran}

\usepackage[T1]{fontenc} 
\usepackage{tikz}
\usepackage{amssymb}
\usepackage{amsthm}
\usepackage{amsmath}
\usepackage{balance}
\usepackage{pifont}
\usepackage{enumitem}
\usepackage{booktabs}
\usepackage{multirow}
\usepackage{pgfplots}
\pgfplotsset{compat=newest}
\usetikzlibrary{intersections}
\usepgfplotslibrary{fillbetween}
\usepackage{graphicx}
\usepackage{glossaries}
\usepackage{array}
\usepackage{comment}
\usepackage{multirow}
\usepackage{makecell}
\usepackage{cite}
\usetikzlibrary{shapes}
\usepackage{float}
\usepackage{stfloats}
\usepackage[capitalise]{cleveref}
\usepackage{verbatim}
\usepackage{textcomp}
\usepackage{titlesec}

\newcommand\mytextsf{\bfseries\sffamily\fontsize{9pt}{9pt}\selectfont}

\newtheorem{theorem}{Theorem}
\newtheorem{proposition}{Proposition}

\newtheorem{lemma}[theorem]{Lemma}

\newtheoremstyle{myremark}
  {\topsep}
  {\topsep}
  {\itshape}
  {0pt}
  {\scshape}
  {.}
  { }
  {}

\theoremstyle{myremark}
\newtheorem{remark}{Remark}

\makeatletter
\let\save@mathaccent\mathaccent
\newcommand*\if@single[3]{%
  \setbox0\hbox{${\mathaccent"0362{#1}}^H$}%
  \setbox2\hbox{${\mathaccent"0362{\kern0pt#1}}^H$}%
  \ifdim\ht0=\ht2 #3\else #2\fi
  }
\newcommand*\rel@kern[1]{\kern#1\dimexpr\macc@kerna}
\newcommand*\widebar[1]{\@ifnextchar^{{\wide@bar{#1}{0}}}{\wide@bar{#1}{1}}}
\newcommand*\wide@bar[2]{\if@single{#1}{\wide@bar@{#1}{#2}{1}}{\wide@bar@{#1}{#2}{2}}}
\newcommand*\wide@bar@[3]{%
  \begingroup
  \def\mathaccent##1##2{%
    \let\mathaccent\save@mathaccent
    \if#32 \let\macc@nucleus\first@char \fi
    \setbox\z@\hbox{$\macc@style{\macc@nucleus}_{}$}%
    \setbox\tw@\hbox{$\macc@style{\macc@nucleus}{}_{}$}%
    \dimen@\wd\tw@
    \advance\dimen@-\wd\z@
    \divide\dimen@ 3
    \@tempdima\wd\tw@
    \advance\@tempdima-\scriptspace
    \divide\@tempdima 10
    \advance\dimen@-\@tempdima
    \ifdim\dimen@>\z@ \dimen@0pt\fi
    \rel@kern{0.6}\kern-\dimen@
    \if#31
      \overline{\rel@kern{-0.6}\kern\dimen@\macc@nucleus\rel@kern{0.4}\kern\dimen@}%
      \advance\dimen@0.4\dimexpr\macc@kerna
      \let\final@kern#2%
      \ifdim\dimen@<\z@ \let\final@kern1\fi
      \if\final@kern1 \kern-\dimen@\fi
    \else
      \overline{\rel@kern{-0.6}\kern\dimen@#1}%
    \fi
  }%
  \macc@depth\@ne
  \let\math@bgroup\@empty \let\math@egroup\macc@set@skewchar
  \mathsurround\z@ \frozen@everymath{\mathgroup\macc@group\relax}%
  \macc@set@skewchar\relax
  \let\mathaccentV\macc@nested@a
  \if#31
    \macc@nested@a\relax111{#1}%
  \else
    \def\gobble@till@marker##1\endmarker{}%
    \futurelet\first@char\gobble@till@marker#1\endmarker
    \ifcat\noexpand\first@char A\else
      \def\first@char{}%
    \fi
    \macc@nested@a\relax111{\first@char}%
  \fi
  \endgroup
}
\makeatother

\makeglossaries

\newacronym{ISAC}{ISAC}{integrated sensing and commmunications}
\newacronym{BS}{BS}{base station}
\newacronym{RF}{RF}{radio-frequency}
\newacronym{DAC}{DAC}{digital-to-analog converter}
\newacronym{IRS}{IRS}{intelligent reflecting surface}
\newacronym{PAPR}{PAPR}{peak-to-average power ratio}

\newacronym{AWGN}{AWGN}{additive white Gaussian noise}
\newacronym{SNR}{SNR}{signal-to-noise ratio}
\newacronym{SINR}{SINR}{signal-to-interference-plus-noise ratio}
\newacronym{SDR}{SDR}{semidefinite relaxation}
\newacronym{SDP}{SDP}{semidefinite programming}
\newacronym{SCA}{SCA}{successive convex approximation}
 
\newacronym{MILP}{MILP}{mixed-integer linear program} 
\newacronym{MINLP}{MINLP}{mixed-integer nonlinear program} 

\newacronym{AoA}{AoA}{angle of arrival}
\newacronym{AoD}{AoD}{angle of departure}
\newacronym{BME}{BME}{beampattern matching error}
 \newacronym{DPG}{DPG}{directional power gain}
 
\newacronym{QoS}{QoS}{quality-of-service} 
\newacronym{LoS}{LoS}{line-of-sight}
\newacronym{NLoS}{NLoS}{non-LoS}

\newacronym{LHS}{LHS}{left-hand-side}
\newacronym{RHS}{RHS}{right-hand-side}
\newacronym{THz}{THz}{terahertz}

\newacronym{BGM}{BGM}{bipartite graph matching}

\newacronym{ES}{ES}{exhaustive search}
\newacronym{BnC}{BnC}{branch-and-cut}

\newacronym{CRB}{CRB}{Cramér-Rao bound}

\begin{document}



\title{\huge Optimal User and Target Scheduling, User-Target Pairing, and Low-Resolution Phase-Only Beamforming for ISAC Systems}
%
%

\author{Luis F. Abanto-Leon and Setareh Maghsudi
\thanks{Copyright (c) 2025 IEEE. Personal use of this material is permitted. However, permission to use this material for any other purposes must be obtained from the IEEE by sending a request to pubs-permissions@ieee.org.}
\thanks{The authors are affiliated with Ruhr University Bochum, Germany.}%
\thanks{The research was supported by the German Federal Ministry of Education and Research under projects 16KISK035 and 16KISK037.}%
}


\maketitle


\begin{abstract}

We investigate the joint user and target scheduling, user-target pairing, and low-resolution phase-only beamforming design for \gls{ISAC}. Scheduling determines which users and targets are served, while pairing specifies which users and targets are grouped into pairs. Additionally, the beamformers are designed using few-bit constant-modulus phase shifts. This resource allocation problem is a nonconvex \gls{MINLP} and challenging to solve. To address it, we propose an exact \gls{MILP} reformulation, which leads to a globally optimal solution. Our results demonstrate the superiority of an optimal joint design compared to heuristic stage-wise approaches, which are highly sensitive to scenario characteristics.



\end{abstract}


\IEEEpeerreviewmaketitle


\glsresetall
\section{Introduction} \label{sec:introduction}



\Gls{ISAC} is an innovative technology offering several advantages, including improved radio resource utilization, reduced costs, and simplified system complexity \cite{liu2020:joint-radar-communication-design-applications-state-art-road-ahead}. However, \gls{ISAC} also introduces new challenges, particularly in designing the radio resource allocation to jointly fulfill users' and targets' requirements.

To enhance sensing accuracy, recent \gls{ISAC} literature has focused on high frequencies, such as millimeter-wave and terahertz bands, using beamforming to mitigate the severe path loss \cite{mao2022:waveform-design-joint-sensing-communications-millimeter-wave-low-terahertz-bands}. However, the high costs of radio components for controlling signals' amplitude and phase at these frequencies have led to the adoption of phase-only beamformers as a more cost-effective solution \cite{dutta2020:case-digital-beamforming-mmwave}. Despite the need for practical solutions, the literature primarily features designs with infinite-resolution phases, e.g., \cite{ lyu2024:crb-minimization-ris-aided-mmwave-integrated-sensing-communications, kaushik2022:waveform-design-joint-radar-communications-low-complexity-analog-components, tsinos2022:dual-function-radar-communication-systems-constant-modulus-similarity-constraints, xiao2024:simultaneous-multi-beam-sweeping-mmwave-massive-mimo-integrated-sensing-communication}, which are infeasible in real-world deployments. Only a few works have accounted for low-resolution phases, e.g., \cite{su2021:secure-dual-functional-radar-communication-transmission-hardware-efficient-design, wang2022:joint-waveform-discrete-phase-shift-design-ris-assisted-integrated-sensing-communication-system-cramer-rao-bound-constraint}, but these designs have relied on approximations to handle phase discretization resulting in inefficient radio resource utilization and highlighting the need for novel approaches that can achieve an optimal design.

Practical systems often cannot simultaneously serve  all users in a system due to limitations, e.g., in the number of \gls{RF} chains, underscoring the need for user scheduling. While this aspect has been well investigated in communication systems, e.g., \cite{abanto2023:radiorchestra-proactive-management-millimeter-wave-self-backhauled-small-cells-joint-optimization-beamforming-user-association-rate-selection-admission-control}, yielding substantial gains, it remains unexplored in \gls{ISAC}. Hence, incorporating user scheduling into \gls{ISAC}'s resource allocation could significantly enhance system performance by enabling optimized decision-making on which users to serve jointly. 

Target scheduling is equally critical, aiming to optimize target selection for sensing while considering resource limitations. While a few studies, e.g., \cite{dou2023:sensing-efficient-noma-aided-integrated-sensing-communication-joint-sensing-scheduling-beamforming-optimization}, have delved into this topic, it remains largely under-explored.



\begin{figure}[!t]
	\centering
	\includegraphics[width=0.92\columnwidth]{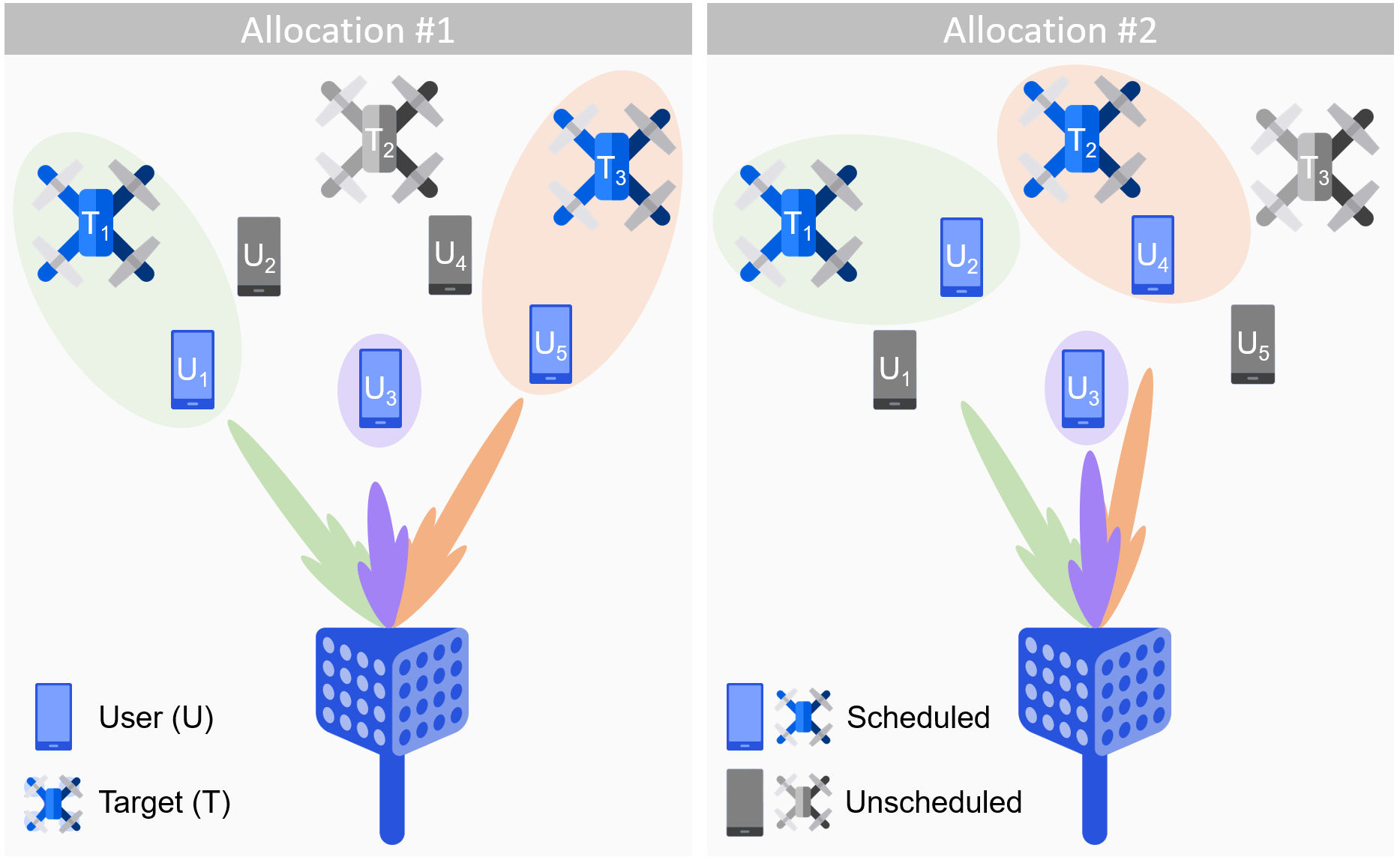}
	\vspace{-2mm}
	\caption{ISAC system consisting of a BS and multiple users and targets. \emph{``Allocation \#1'' shows a favorable resource allocation strategy, featuring well-separated pairs and strong alignment between users and targets within each pair. This ensures high directivity, benefiting both users and targets. The sufficient spacing between scheduled users ($\mathsf{U}_1$, $\mathsf{U}_3$, $\mathsf{U}_5$) minimizes inter-user interference, while the adequate separation between targets ($\mathsf{T}_1$, $\mathsf{T}_3$) reduces cross-interference. In contrast, ``Allocation \#2'' is less optimal due to closer pairs spread over a wider angular range, resulting in weaker alignment and reduced directivity. The angular proximity of users ($\mathsf{U}_2$, $\mathsf{U}_3$, $\mathsf{U}_4$) increases inter-user interference, and the limited separation between targets ($\mathsf{T}_1$, $\mathsf{T}_2$) heightens cross-correlation. While this scenario highlights the importance of alignment, a more comprehensive resource allocation approach is needed. Factors such as user channel conditions, target characteristics, and phase resolution significantly influence performance, necessitating a unified optimization strategy beyond fixed pairing and scheduling criteria.}}	
	\label{fig:system-model}
	\vspace{-0.5cm}
\end{figure}


Most \gls{ISAC} works have assumed that user-target pairing is predetermined and provided as prior information, e.g., \cite{kaushik2022:waveform-design-joint-radar-communications-low-complexity-analog-components, tsinos2022:dual-function-radar-communication-systems-constant-modulus-similarity-constraints, zhao2022:joint-transmit-receive-beamforming-design-integrated-sensing-communication}, which is employed for subsequent resource allocation. Recently, however, there has been a shift towards exploring flexible user-target pairing, e.g., \cite{dou2024:channel-sharing-aided-integrated-sensing-communication-energy-efficient-sensing-scheduling-approach, cazzella2024:deep-learning-based-target-to-user-association-integrated-sensing-communication-systems}, allowing for improved performance by dynamically associating users and targets based on factors such as location and channel conditions. Yet, existing studies on this topic have only considered one user per time slot, e.g., \cite{dou2024:channel-sharing-aided-integrated-sensing-communication-energy-efficient-sensing-scheduling-approach, cazzella2024:deep-learning-based-target-to-user-association-integrated-sensing-communication-systems}, thereby avoiding multiuser interference. Exploiting spatial multiplexing to service multiple users and targets simultaneously is crucial in modern radar and communication systems, as it improves radio resource utilization, making it a key aspect to consider.


Jointly designing scheduling, pairing, and beamforming is a challenging task. Although performing these processes in separate stages can simplify the design, this decoupling often results in suboptimal performance. Particularly, pairing based on a single criterion, such as alignment, as shown in \cref{fig:system-model}, is a valid approach. However, pairing cannot be treated in isolation from scheduling, as both are influenced by users' channel characteristics and targets' features. Furthermore, phase resolution constrains the beamforming directions, further emphasizing the interdependence between scheduling, pairing, and beamforming. This paper investigates the optimal joint design of these three  key processes, resulting in a novel resource allocation problem that is formulated as a nonconvex \gls{MINLP}. We propose a reformulation that transforms the nonconvex \gls{MINLP} into a \gls{MILP}, enabling global optimality through convex transformations that preserve the original solution space. Simulations demonstrate that our joint design provides significant performance gains over four baseline methods inspired by existing literature, which rely on heuristic scheduling and pairing.

\emph{Notation}: Matrices and vectors are respectively denoted by $ \mathbf{A} $ and $ \mathbf{a} $. The transpose, Hermitian transpose, and trace of $ \mathbf{A} $ are denoted by $ \mathbf{A}^\mathrm{T} $, $ \mathbf{A}^\mathrm{H} $, and $ \mathrm{Tr} \left(  \mathbf{A} \right) $, respectively. The $l$-th row and $i$-th column of $ \mathbf{A} $ are denoted by $ \left[ \mathbf{A} \right]_{l,:} $ and $ \left[ \mathbf{A} \right]_{:,i} $, respectively, and the $l$-th element of $ \mathbf{a} $ is denoted by $ \left[ \mathbf{a} \right]_l $. 


\section{System Model and Problem Formulation} \label{sec:system-model-problem-formulation}

We consider a multi-user, multi-target downlink \gls{ISAC} system, as illustrated in \cref{fig:system-model}, where the \gls{BS} jointly performs sensing and communication tasks.

\textbf{Preliminaries:} The \gls{BS} has $ N $ transmit antennas and $ N $ receive antennas, indexed by set $ \mathcal{N} = \left\lbrace 1, \dots, N \right\rbrace $. Also, there are $ T $ targets and $ U $ single-antenna users, indexed by sets $ \mathcal{T} = \left\lbrace 1, \dots, T \right\rbrace $ and $ \mathcal{U} = \left\lbrace 1, \dots, U \right\rbrace $, respectively. The \gls{BS} has $ K $ \gls{RF} chains, where $ K \leq U $. Each \gls{RF} chain supports a single user's data stream.
Following the current industry stance on \gls{ISAC}, which advocates prioritizing communications while  enabling sensing opportunistically, we predestine the use of \gls{RF} chains primarily to serve user demands. Thus, users can be served individually or jointly with a target, but targets are not allocated \gls{RF} chains for sensing only. The \gls{BS} chooses $ J $ targets to sense from the total $ T $, where $ J \leq \min \left\lbrace K, T \right\rbrace $. In particular, each target is sensed simultaneously while servicing exactly one user\footnote{To maintain a practical perspective and prevent performance degradation \cite{dou2024:channel-sharing-aided-integrated-sensing-communication-energy-efficient-sensing-scheduling-approach}, we adopt a one-to-one pairing. Yet, our approach can be extended to a many-to-one pairing, allowing a user to be matched with multiple targets.}.  In the sequel, we denote the $ u $-th user and the $ t $-th target by $ \mathsf{U}_u $ and $ \mathsf{T}_t $, respectively.

\textbf{User scheduling:}
As the \gls{RF} chains are limited, only a subset of users can be scheduled at a given channel use\footnote{We assume that unscheduled users are queued for a subsequent channel use, where resource allocation is re-executed to accommodate users who were not served in the current channel use, as well as any new arriving users requiring service.}. Thus, we introduce constraints  
\begin{align*}
	\mathrm{C}_{1}: & \mu_u \in \left\lbrace 0, 1 \right\rbrace, \forall u \in \mathcal{U},
	\\
	\mathrm{C}_{2}: & \textstyle \sum_{u \in \mathcal{U}} \mu_u = K.
\end{align*}

In $ \mathrm{C}_{1} $, $ \mu_u = 1 $  indicates that $ \mathsf{U}_u $ is scheduled, and $ \mu_u = 0 $ otherwise. Besides, $ \mathrm{C}_{2} $ enforces that all \gls{RF} chains are used for communication, operating the system at maximum capacity.

\textbf{Target scheduling:} 
As the number of targets can be large, sensing all targets simultaneously is not feasible. To determine which targets are sensed in the current channel use, we add
\begin{align*}
	\mathrm{C}_{3}: & \lambda_t \in \left\lbrace 0, 1 \right\rbrace, \forall t \in \mathcal{T},
	\\
	\mathrm{C}_{4}: & \textstyle \sum_{t \in \mathcal{T}} \lambda_t = J.
\end{align*}

In $ \mathrm{C}_{3} $, $ \lambda_t = 1 $ indicates that $ \mathsf{T}_t $ is scheduled, and $ \lambda_t = 0 $ otherwise, while $ \mathrm{C}_{4} $ limits the scheduled targets to $ J $.

\textbf{User-target pairing:}
Some targets are paired with scheduled users, while others remain unpaired and unscheduled in the current channel use. To depict this, we add constraints 
\begin{align*}
	\mathrm{C}_{5}: & \rho_{u,t} \in \left\lbrace 0, 1\right\rbrace, \forall u \in \mathcal{U}, t \in \mathcal{T},
	\\
	\mathrm{C}_{6}: & \textstyle \sum_{u \in \mathcal{U}} \rho_{u,t} = \lambda_t, \forall t \in \mathcal{T},
	\\
	\mathrm{C}_{7}: & \textstyle \sum_{t \in \mathcal{T}} \rho_{u,t} \leq \mu_u, \forall u \in \mathcal{U}.
\end{align*}

In $ \mathrm{C}_{5} $, $ \rho_{u,t} = 1 $ indicates that $ \mathsf{U}_u $ and $ \mathsf{T}_t $ are paired, and $ \rho_{u,t} = 0 $ otherwise. Also, $ \mathrm{C}_{6} $ ensures that a scheduled target is paired with only one user, while $ \mathrm{C}_{7} $ restricts pairing to scheduled users only as sensing piggybacks on communications.

\textbf{Communication model:}
The \gls{BS} transmits signal $ \mathbf{d} = \textstyle \sum_{u \in \mathcal{U}} \mathbf{w}_u s_u $, where $ \mathbf{w}_u \in \mathbb{C}^{N \times 1} $ is the beamforming vector for $ \mathsf{U}_u $, and $ s_u \in \mathbb{C} $ is the data symbol for $ \mathsf{U}_u $, which follows
a complex Gaussian distribution with zero mean and unit variance, e.g., $ \mathbb{E} \left\lbrace s_u s_u^{*} \right\rbrace = 1 $. The signal received by $ \mathsf{U}_u $ is $ y_u = \textstyle \mathbf{h}^\mathrm{H}_u \mathbf{d} + n_u = \sum_{u \in \mathcal{U}} \mathbf{h}^\mathrm{H}_u \mathbf{w}_u s_u + n_u  $, where $ \mathbf{h}_u \in \mathbb{C}^{N \times 1} $ is the channel between the \gls{BS} and $ \mathsf{U}_u $, and $ n_u \sim \mathcal{CN} \left( 0,\sigma^2 \right) $ is \gls{AWGN}. The \gls{SINR} of $ \mathsf{U}_u $ is 
\begin{equation}
	\mathsf{SINR}_u \left( \mathbf{W} \right) = \big| \widetilde{\mathbf{h}}^\mathrm{H}_u \mathbf{w}_u \big|^2 / \big( \textstyle \sum_{i \in \mathcal{U}, i \neq u} \big| \widetilde{\mathbf{h}}^\mathrm{H}_u \mathbf{w}_i\big|^2  + 1 \big),
\end{equation}
where $ \widetilde{\mathbf{h}}_u = \frac{ \mathbf{h}_u }{\sigma} $ and $ \mathbf{W} = \left[ \mathbf{w}_1, \dots, \mathbf{w}_U \right] $. Now, we add 
\begin{equation} \nonumber
	\mathrm{C}_{8}:
		\begin{cases}
			   	\mathsf{SINR}_u \left( \mathbf{W} \right) \geq \Gamma_{\mathrm{th}, u}, & \text{if} ~ \mu_u = 1
			   	\\	
				\mathsf{SINR}_u \left( \mathbf{W} \right) = 0, & \text{if} ~ \mu_u = 0
		\end{cases}, \forall u \in \mathcal{U},
\end{equation}
enforcing a \gls{SINR} threshold $ \Gamma_{\mathrm{th},u} $ on scheduled users $ \left( \mu_u = 1 \right) $ while not enforcing it on unscheduled users $ \left( \mu_u = 0 \right) $.

\textbf{Sensing model:} The \gls{BS} operates as a monostatic co-located radar, i.e., the \gls{AoD} and \gls{AoA} are identical. The targets are modeled as single points, assuming they are far from the \gls{BS}. The \gls{BS} transmits signals in the directions of the targets, using the response matrices. The response matrix between the \gls{BS} and $ \mathsf{T}_t $ is 
$
	\mathbf{G}_t = \alpha_t \mathbf{a} \left( \theta_t \right) \mathbf{a}^\mathrm{H} \left( \theta_t \right), \forall t \in \mathcal{T}
$,
where $ \alpha_t $ is the reflection coefficient of $ \mathsf{T}_t $ \cite{li2007:mimo-radar-colocated-antennas}, $ \theta_t $ is the \gls{AoD}/\gls{AoA} of $ \mathsf{T}_t $, and $ \mathbf{a} \left( \theta \right) = \left[ e^{j \pi \frac{-N+1}{2} \cos \left( \theta \right)}, \dots, e^{j \pi \frac{N-1}{2}\cos \left( \theta \right)} \right]^\mathrm{T} \in \mathbb{C}^{N \times 1} $ is the steering vector in the direction of $ \theta $, assuming half-wavelength antenna spacing. The \gls{DPG} is adopted as a design criterion for sensing \cite{ li2023:beamforming-design-active-irs-aided-mimo-integrated-sensing-communication-systems}, given by
\begin{equation}
	\mathsf{DPG}_t \left( \mathbf{v}_t \right) = \mathbf{v}^\mathrm{H}_t \mathbf{G}_t \mathbf{v}_t, \forall t \in \mathcal{T},
\end{equation}
where $ \mathbf{v}_t $ is the beamforming vector used for illuminating $ \mathsf{T}_t $. Improving the \gls{DPG} is crucial, as it increases the power radiated towards the targets, thereby enhancing detectability \cite{li2023:beamforming-design-active-irs-aided-mimo-integrated-sensing-communication-systems, meng2022:intelligent-reflecting-surface-enabled-multi-target-sensing}. Hence, we first introduce constraint
\begin{align*}
	\mathrm{C}_{9}: \tau > 0,
\end{align*}
where $ \tau $ is an auxiliary variable, and then include constraint
\begin{align*}
	\mathrm{C}_{10}: 
		\begin{cases}
			   	\mathbf{v}^\mathrm{H}_t \mathbf{G}_t \mathbf{v}_t \geq \tau, & \text{if} ~ \lambda_t = 1
			   	\\	
				\mathbf{v}^\mathrm{H}_t \mathbf{G}_t \mathbf{v}_t = 0, & \text{if} ~ \lambda_t = 0
		\end{cases}, \forall t, \in \mathcal{T}, 
\end{align*}
to bound the \gls{DPG} of the scheduled targets. We also add 
\begin{align*}
	\resizebox{0.98\columnwidth}{!}{$
	\mathrm{C}_{11}: 
		\begin{cases}
			   	\mathbf{v}^\mathrm{H}_t \mathbf{G}_q \mathbf{v}_t \leq \xi_\mathrm{th}, & \text{if} ~ \lambda_t = 1 ~\text{and}~ \lambda_q = 1
			   	\\	
				\mathbf{v}^\mathrm{H}_t \mathbf{G}_q \mathbf{v}_t < \infty, & \text{otherwise}
		\end{cases}, \forall t,q \in \mathcal{T}, t \neq q,
		$}
\end{align*}
to limit the cross-interference power among scheduled targets, where $ \xi_\mathrm{th} $ is the maximum acceptable threshold \cite{meng2022:intelligent-reflecting-surface-enabled-multi-target-sensing}.

\textbf{Phase-only beamforming:}
The beamforming is designed with low-resolution constant-modulus discrete phases, given by set $ \mathcal{S} = \left\lbrace  \delta e^{j \phi_1}, \dots, \delta e^{j \phi_L} \right\rbrace $, where $ \phi_l $ is the $ l $-th phase, $ \delta = \sqrt{\tfrac{P_\mathrm{tx}}{K N}} $ is the magnitude, $ Q $ is the number of bits needed for representing the $ L $ phases in $ \mathcal{S} $, and $ P_\mathrm{tx} $ is the \gls{BS}'s transmit power. Thus, we include constraint
\begin{equation} \nonumber
	\mathrm{C}_{12}:
		\begin{cases}
			   	\left[ \mathbf{w}_u \right]_n \in \mathcal{S}, & \text{if} ~ \mu_u = 1
			   	\\	
				\left[ \mathbf{w}_u \right]_n = 0, & \text{if} ~ \mu_u = 0
		\end{cases}, \forall u \in \mathcal{U}, n \in \mathcal{N},
\end{equation}
which represents the beamforming design criteria for scheduled $ \left( \mu_u = 1 \right) $ and unscheduled $ \left( \mu_u = 0 \right) $ users. Since targets are paired and co-scheduled with users, then $ \mathbf{v}_t $ must be equal to one of the beamforming vectors $ \mathbf{w}_u $, which is ensured by
\begin{align*}
	\mathrm{C}_{13}: \mathbf{v}_t = \textstyle \sum_{u \in \mathcal{U}} \mathbf{w}_u \rho_{u,t}, \forall t \in \mathcal{T}.
\end{align*}

\textbf{Problem formulation:}
The joint user and target scheduling, user-target pairing, and low-resolution beamforming design is formulated by problem $ \mathcal{P} $, whose objective is to maximize the minimum \gls{DPG} of all scheduled targets via the use of $ \tau $.
\begin{align*} 
	\mathcal{P}: ~~ \underset{ \boldsymbol{\Omega}_\mathcal{P} }{\mathrm{maximize}} ~~ \textstyle \tau ~~ \mathrm{s.t.} ~~ \mathrm{C}_{1} - \mathrm{C}_{13}. 
\end{align*}

Set $ \boldsymbol{\Omega}_\mathcal{P} $ encompasses all decision variables of $ \mathcal{P} $, specifically, $ \tau $, $ \mu_u $, $ \lambda_t $, $ \rho_{u,t} $, $ \mathbf{w}_u $, and $ \mathbf{v}_t $. In particular, $ \mathcal{P} $ is a nonconvex \gls{MINLP}, making it challenging to solve. 


\section{Proposed Optimal Approach} \label{sec:proposed-approach}


We propose a series of equivalent transformations, detailed in \cref{thm:proposition-1} to \cref{thm:proposition-8}, to reformulate the nonconvex \gls{MINLP} $ \mathcal{P} $ into a convex, linear \gls{MILP} $ \mathcal{Q} $. This reformulation uncovers hidden convexities within $ \mathcal{P} $, enabling its transformation into a tractable form without altering the original solution space. By preserving the original solution space at each step of the reformulation, we guarantee that an optimal solution to problem $ \mathcal{Q} $ is also optimal to problem $ \mathcal{P} $. \emph{The proofs for all subsequent propositions are provided in the Appendix.}

\begin{proposition} \label{thm:proposition-1}
	Constraint $ \mathrm{C}_{12} $ can be equivalently rewritten as constraints $ \mathrm{D}_{1} $, $ \mathrm{D}_{2} $, and $ \mathrm{D}_{3} $, 
	\begin{equation} \nonumber
		\mathrm{C}_{12} \Leftrightarrow
			\begin{cases}
				   	\mathrm{D}_{1}: \left[ \mathbf{x}_{u,n} \right]_l \in \left\lbrace 0, 1\right\rbrace, \forall u \in \mathcal{U}, n \in \mathcal{N}, l \in \mathcal{L}, 
				   	\\	
				   	\mathrm{D}_{2}: \mathbf{1}^\mathrm{T} \mathbf{x}_{u,n} = \mu_u, \forall u \in \mathcal{U}, n \in \mathcal{N},  
				   	\\
				   	\mathrm{D}_{3}: \left[ \mathbf{w}_u \right]_n = \mathbf{s}^\mathrm{T} \mathbf{x}_{u,n}, \forall u \in \mathcal{U}, n \in \mathcal{N}, 
			\end{cases}
	\end{equation}
	where vector $ \mathbf{s} \in \mathbb{C}^{L \times 1} $ is formed by the elements in $ \mathcal{S} $ and $ \mathcal{L} = \left\lbrace 1, \dots, L \right\rbrace $.
\end{proposition}

\begin{figure*}[!t]
	\begin{equation} \nonumber
		\mathrm{C}_{7} \Leftrightarrow
			\begin{cases}
				   	\mathrm{E}_{1}: \mathbf{W}_u = \mathbf{w}_u \mathbf{w}^\mathrm{H}_u, \forall u \in \mathcal{U}, 
				   	~~~
				   	\mathrm{E}_{2}: \sum_{i \in \mathcal{U}} \mathrm{Tr} \big( \widetilde{\mathbf{H}}_u \mathbf{W}_i \big) + 1 \leq \big( 1 + \Gamma_{\mathrm{th},u}^{-1} \big) \mathrm{Tr} \big( \widetilde{\mathbf{H}}_u \mathbf{W}_u \big) + \big( 1 - \mu_u \big) \widebar{B}_u, \forall u \in \mathcal{U}, 
			\end{cases}
	\end{equation}
	\hrulefill
	\begin{equation} \nonumber
	\resizebox{1.02\textwidth}{!}{$
		\mathrm{C}_{11}, \mathrm{C}_{13} 
		\Leftrightarrow
			\begin{cases}
			\mathrm{K}_\mathrm{1}: \pi_{t,q} \leq \lambda_t, \forall t, q \in \mathcal{T}, t \neq q,
			~~~~~~
			\mathrm{K}_\mathrm{2}: \pi_{t,q} \leq \lambda_q, \forall t, q \in \mathcal{T}, t \neq q,
			~~~~~~
			\mathrm{K}_\mathrm{3}: \pi_{t,q} \geq \lambda_t + \lambda_q - 1, \forall t, q \in \mathcal{T}, t \neq q,
			\\
			\mathrm{K}_\mathrm{4}: \pi_{t,q} \in \left[ 0, 1 \right], \forall t, q \in \mathcal{T}, t \neq q,
			~~~~~~
			\mathrm{K}_\mathrm{5}: \mathrm{Tr} \left( \mathbf{G}_q \mathbf{W}_u  \right) \leq \xi_\mathrm{th} + \left( 2 - \pi_{t,q} - \rho_{u,t} \right) \widebar{D}_t, \forall u \in \mathcal{U}, t,q \in \mathcal{T}, t \neq q,
			\end{cases}	
			$}
	\end{equation}
	\hrulefill
\end{figure*}

\begin{proposition} \label{thm:proposition-2}
	Constraint $ \mathrm{C}_{8} $ can be equivalently rewritten as constraints $ \mathrm{E}_{1} $ and $ \mathrm{E}_{2} $ (at the top of next page), where $ \widetilde{\mathbf{H}}_u = \widetilde{\mathbf{h}}_u \widetilde{\mathbf{h}}^\mathrm{H}_u $ and $ B_\mathrm{max} = P_\mathrm{tx} \cdot \mathrm{Tr} \big( \widetilde{\mathbf{H}}_u \big) + 1 $ is an upper bound for the \gls{LHS} of $ \mathrm{E}_{2} $.
\end{proposition}

\begin{proposition} \label{thm:proposition-3}
	Constraint $ \mathrm{E}_{1} $ can be equivalently rewritten as constraint $ \mathrm{F}_{1} $, 
	\begin{equation} \nonumber
		\mathrm{E}_{1} \Leftrightarrow \mathrm{F}_{1}: \left[ \mathbf{W}_u \right]_{n,m} = \left[ \mathbf{w}_u \right]_n \left[ \mathbf{w}^{*}_u \right]_m, \forall u \in \mathcal{U}, n, m \in \mathcal{N},
	\end{equation}
	where $ \left[ \mathbf{W}_u \right]_{n,m} $ represents the element of $ \mathbf{W}_u $ in the $ n $-th row and $ m $-th column.
\end{proposition}

\begin{proposition} \label{thm:proposition-4}
	Constraints $ \mathrm{D}_{3} $ and $ \mathrm{F}_{1} $ can be equivalently expressed as constraints $ \mathrm{G}_{1} $, $ \mathrm{G}_{2} $, and $ \mathrm{G}_{3} $, 
	\begin{equation} \nonumber
		\resizebox{1.03\columnwidth}{!}{$
		\begin{matrix}
		\mathrm{D}_{3} \\
		\mathrm{F}_{1} 
		\end{matrix}
		\Leftrightarrow
			\begin{cases}
				   	\mathrm{G}_{1}: \left[ \mathbf{W}_u \right]_{n,m} = \mathrm{Tr} \left( \mathbf{S} \mathbf{x}_{u,n} \mathbf{x}^\mathrm{T}_{u,m} \right), \forall u \in \mathcal{U}, n \in \mathcal{N}, m \in \mathcal{M}_n,
				   	\\	
				   	\mathrm{G}_{2}: \left[ \mathbf{W}_u \right]_{m,n} = \left[ \mathbf{W}_u \right]^{*}_{n,m}, \forall u \in \mathcal{U}, n \in \mathcal{N}, m \in \mathcal{M}_n,
				   	\\	
				   	\mathrm{G}_{3}: \left[ \mathbf{W}_u \right]_{n,n} = \delta^2 \mu_u, \forall u \in \mathcal{U}, n \in \mathcal{N},  
			\end{cases}
			$}
	\end{equation}
	where $ \mathbf{S} = \mathbf{s}^{*} \mathbf{s}^\mathrm{T} $ and $ \mathcal{M}_n = \left\lbrace n+1, \dots, N \right\rbrace $.
\end{proposition}

\begin{proposition} \label{thm:proposition-5}
	Constraint $ \mathrm{G}_{1} $ can be equivalently expressed as constraints $ \mathrm{H}_{1} $, $ \mathrm{H}_{2} $, and $ \mathrm{H}_{3} $, 
	\begin{equation} \nonumber
		\resizebox{1.03\columnwidth}{!}{$
		\mathrm{G}_{1} \Leftrightarrow
			\begin{cases}
					\mathrm{H}_{1}: \mathbf{Y}_{u, n,m} = \mathbf{x}_{u,n} \mathbf{x}^T_{u,m}, \forall u \in \mathcal{U}, n \in \mathcal{N}, m \in \mathcal{M}_n,
					\\
					\mathrm{H}_{2}: \left[ \mathbf{W}_u \right]_{n,m} = \mathrm{Tr} \left( \mathbf{S} \mathbf{Y}_{u,n,m}  \right), \forall u \in \mathcal{U}, n \in \mathcal{N}, m \in \mathcal{M}_n,  
					\\
					\mathrm{H}_{3}: \left[ \mathbf{Y}_{u, n,m} \right]_{l,i} \in \left\lbrace 0, 1\right\rbrace, \forall u \in \mathcal{U}, n \in \mathcal{N}, m \in \mathcal{M}_n, l, i \in \mathcal{L}, 
			\end{cases}
			$}
	\end{equation}
\end{proposition}

\begin{proposition} \label{thm:proposition-6}
	Constraints $ \mathrm{H}_{1} $ and $ \mathrm{H}_{3} $ can be equivalently expressed as constraints $ \mathrm{I}_{1} $, $ \mathrm{I}_{2} $, and $ \mathrm{I}_{3} $,
	\begin{equation} \nonumber
	\resizebox{1.03\columnwidth}{!}{$
		\begin{matrix}
		\mathrm{H}_{1} \\
		\mathrm{H}_{3} 
		\end{matrix}
		\Leftrightarrow
			\begin{cases}
				   	\mathrm{I}_{1}: \mathbf{1}^\mathrm{T} \left[ \mathbf{Y}_{u,n,m} \right]_{:,i} = \left[ \mathbf{x}_{u,m} \right]_{i}, \forall u \in \mathcal{U}, n \in \mathcal{N}, m \in \mathcal{M}_n, i \in \mathcal{L},
				   	\\	
				   	\mathrm{I}_{2}: \left[ \mathbf{Y}_{u,n,m} \right]_{l,:} \mathbf{1} = \left[ \mathbf{x}_{u,n} \right]_l, \forall u \in \mathcal{U}, n \in \mathcal{N}, m \in \mathcal{M}_n, l \in \mathcal{L}, 
				   	\\
				   	\mathrm{I}_{3}: \left[ \mathbf{Y}_{u, n,m} \right]_{l,i} \in \left[ 0, 1 \right], \forall u \in \mathcal{U}, n \in \mathcal{N}, m \in \mathcal{M}_n, l, i \in \mathcal{L}, 
			\end{cases}
			$}
	\end{equation}
\end{proposition}

\begin{proposition} \label{thm:proposition-7}
	Constraint $ \mathrm{C}_{10} $ is equivalent to constraint $ \mathrm{J}_{1} $,
	\begin{equation*}
	\resizebox{1.0\columnwidth}{!}{$
		\mathrm{C}_{10}
		\Leftrightarrow
			\mathrm{J}_\mathrm{1}: \mathrm{Tr} \left( \mathbf{G}_t \mathbf{W}_u  \right) \geq \tau - \left( 2 - \lambda_t - \rho_{u,t} \right) \widebar{D}_t, \forall u \in \mathcal{U}, t \in \mathcal{T},
			$}
	\end{equation*}
	where $ \widebar{D}_t = \tfrac{P_\mathrm{tx}}{K} \cdot \mathrm{Tr} \big( \mathbf{G}_t \big) $.
\end{proposition}

\begin{proposition} \label{thm:proposition-8}
	Constraints $ \mathrm{C}_{11} $ and $ \mathrm{C}_{13} $ are equivalent to constraints $ \mathrm{K}_{1} $, $ \mathrm{K}_{2} $, $ \mathrm{K}_{3} $, $ \mathrm{K}_{4} $, and $ \mathrm{K}_{5} $ (at the top of next page).
\end{proposition}

After applying the propositions above, nonconvex \gls{MINLP} $ \mathcal{P} $ is equivalently recast as
\begin{align*} 
	\mathcal{Q}: & ~ \underset{\boldsymbol{\Omega}_\mathcal{Q} }{\mathrm{maximize}}  
	& & \tau
	\\
	& ~~~~~ \mathrm{s.t.} & & \mathrm{C}_{1}, \mathrm{C}_{2}, \mathrm{C}_{3}, \mathrm{C}_{4}, \mathrm{C}_{5}, \mathrm{C}_{6}, \mathrm{C}_{7}, \mathrm{C}_{9}, \mathrm{D}_{1}, \mathrm{D}_{2}, \mathrm{E}_{2}, 
	\\
	& & & \mathrm{G}_{2}, \mathrm{G}_{3}, \mathrm{H}_{2}, \mathrm{I}_{1}, \mathrm{I}_{2}, \mathrm{I}_{3}, \mathrm{J}_{1}, \mathrm{K}_{1}, \mathrm{K}_{2}, \mathrm{K}_{3}, \mathrm{K}_{4}, \mathrm{K}_{5}, 
\end{align*}
where  $ \boldsymbol{\Omega}_\mathcal{Q} $ represents all decision variables of $ \mathcal{Q} $, which include $ \tau $, $ \mu_u $, $ \lambda_t $, $ \rho_{u,t} $, $ \pi_{t,q} $, $ \mathbf{w}_u $, $ \mathbf{x}_{u,n} $, $ \mathbf{W}_u $, and $ \mathbf{Y}_{u,n,m} $. In particular, $ \mathcal{Q} $ is an \gls{MILP}, a convex optimization problem that can be solved to global optimality.

\begin{remark} 
	The worst-case computational complexity of $ \mathcal{P} $ is an \gls{ES} requiring the evaluation of $ \mathcal{C}_\mathrm{ES} = 2^{K Q N} {U \choose K} T! $ candidate solutions. Nevertheless, the special structure of $ \mathcal{Q} $ allows us to utilize \gls{BnC}, implemented in commercial solvers, which can solve $ \mathcal{Q} $ optimally at a small fraction of $ \mathcal{C}_\mathrm{ES} $. In particular, \gls{BnC} operates by pruning suboptimal and infeasible candidate solutions \cite{desrosiers2010:branch-price-cut-algorithms}, which lies beyond the scope of this work.
\end{remark}

\section{Simulation Results} \label{sec:simulation-results}

We evaluate our proposed approach using the Rician fading channel model. Thus, the channel for $ \mathsf{U}_u $ is given by $ \mathbf{h}_u = \gamma_u \mathbf{v}_u $, where $ \gamma_u $ accounts for large-scale fading and $ \mathbf{v}_u = 1 / \sqrt{K_\mathrm{R}+1} \left( \sqrt{K_\mathrm{R}} \cdot \mathbf{v}^\mathrm{LoS}_u + \mathbf{v}^\mathrm{NLoS}_u \right) $ is the normalized small-scale fading, with $ K_\mathrm{R} $ denoting the fading factor. The  \gls{LoS} component is $ \mathbf{v}^\mathrm{LoS}_u = \mathbf{a} \left( \beta_u \right) $, where $ \beta_u $ is the \gls{LoS} angle, and the \gls{NLoS} components are $ \mathbf{v}^\mathrm{NLoS}_u \sim \mathcal{CN} \left( \mathbf{0}, \mathbf{I} \right) $. The \gls{LoS} components are randomized but maintaining a separation of $ \Delta = \beta_{u} - \beta_{u-1} $ between them to control the level of channel correlation. Here, $ \gamma_u = 28 + 22 \log_{10} (d_u) + 20 \log_{10} (f_\mathrm{c}) + \chi_u $ is computed using the UMa model \cite{3gpp:38.901}, where $ f_\mathrm{c} $ is the carrier frequency, $ \chi_u \sim \mathcal{CN} \left( 0, \zeta \right) $ is the shadowing for $ \mathsf{U}_u $, and $ d_u $ is the distance between the \gls{BS} and $ \mathsf{U}_u $. We simulate $ 100 $ realizations considering deterministic parameters $ K_\mathrm{R} = 100 $, $ U = 5 $, $ K = 2 $, $ T = 4 $, $ J = 2 $, $ N = 12 $ (cf. \cite{sibeam2016:datasheet}), $ \xi_\mathrm{th} = 0.01 $, $ Q = 2 $, $ \mathcal{S} = \left\lbrace \delta, j \delta, -\delta, -j \delta \right\rbrace $, $ f_\mathrm{c} = 71 $ GHz, $ \sigma^2 = -87 $ dBm, $ \zeta = 4 $ dB, and random parameters $ \alpha_t \in \left[ 0.04, 0.08 \right] $, $ \theta_t \in \left[ 20, 160 \right] $, and $ d_u = \left[ 20, 80 \right] $m. We benchmark the following approaches, solved using \textsf{CVX} and \textsf{MOSEK}, with a tolerance of $ 0.01 \% $, below which the solution is assumed to be optimal.

\textbf{\textsf{Optimal scheduling, pairing, and beamforming} }({\mytextsf{OPT-SPB}}): This is the proposed approach, which yields an optimal design.

\textbf{\textsf{Baseline 1 }}({\mytextsf{BL1}}): The pairing and scheduling are heuristic, while the beamforming is computed optimally as in {\mytextsf{OPT-SPB}}. This baseline prioritizes pairing over scheduling. Inspired by \cite{dou2024:channel-sharing-aided-integrated-sensing-communication-energy-efficient-sensing-scheduling-approach}, the pairing is performed via \gls{BGM}, maximizing the sum of edge weights and yielding $ T $ user-target pairs. Each edge weight $ \omega_{u,t} = \tfrac{ \left| \mathbf{h}^\mathrm{H}_u \mathbf{a} \left( \theta_t \right) \right| }{\left| \mathbf{h}_u \right| \left| \mathbf{a} \left( \theta_t \right) \right| } $ quantifies the alignment between $ \mathsf{U}_u $ and $ \mathsf{T}_t $.  

\textbf{\textsf{Baseline 2 }}({\mytextsf{BL2}}): The pairing and scheduling are heuristic, while the beamforming is computed optimally as in {\mytextsf{OPT-SPB}}.  This baseline prioritizes scheduling over pairing. Inspired by \cite{dou2024:channel-sharing-aided-integrated-sensing-communication-energy-efficient-sensing-scheduling-approach, christopoulos2015:multicast-multigroup-precoding-user-scheduling-frame-based-satellite-communications}, the scheduling involves choosing $ K $ users with the least mutual channel correlation. The correlation is defined as $ \xi_{u,i} = \tfrac{\left| \mathbf{h}^\mathrm{H}_u \mathbf{h}_i \right|}{\left| \mathbf{h}_u \right| \left| \mathbf{h}_i \right|} $ for any two users $ \mathsf{U}_u $ and $ \mathsf{U}_{i \neq u} $. The pairing is done via \gls{BGM} as in {\mytextsf{BL1}}, but using only the scheduled users.

\textbf{\textsf{Baseline 3 }}({\mytextsf{BL3}}): Scheduling is random, choosing $ K $ users from $ U $. Pairing and beamforming are optimal as in {\mytextsf{OPT-SPB}}.

\textbf{\textsf{Baseline 4 }}({\mytextsf{BL4}}): Scheduling and pairing are performed randomly. The beamforming is optimal as in {\mytextsf{OPT-SPB}}.

\begin{figure}[!t]
	\centering
	\includegraphics[width=0.96\columnwidth]{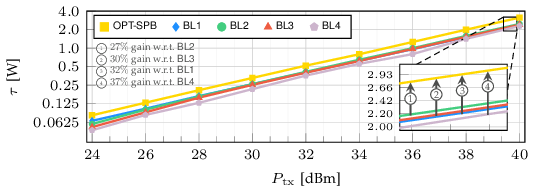}
	\vspace{-3.5mm}
	\caption{Performance as a function of transmit power.}	
	\label{fig:scenario-1}
	\vspace{-0.75mm}
\end{figure}
\begin{figure}[!t]
	\centering
	\includegraphics[width=0.96\columnwidth]{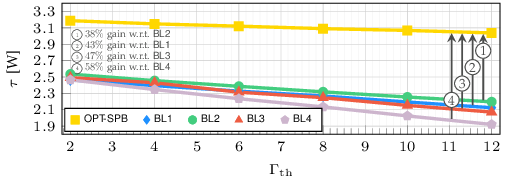}
	\vspace{-3.5mm}
	\caption{Performance as a function of SINR threshold in correlated channels.}	
	\label{fig:scenario-2}
	\vspace{-0.75mm}
\end{figure}
\begin{figure}[!t]
	\centering
	\includegraphics[width=0.96\columnwidth]{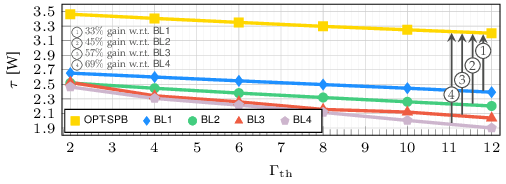}
	\vspace{-3.5mm}
	\caption{Performance as a function of SINR threshold in uncorrelated channels.}	
	\label{fig:scenario-3}
	\vspace{-0.75mm}
\end{figure}
\begin{figure}[!t]
	\centering
	\includegraphics[width=0.96\columnwidth]{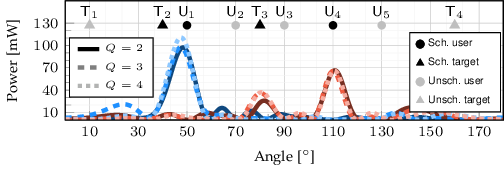}
	\vspace{-3.5mm}
	\caption{Beampatterns for scheduled and paired users and targets.}	
	\label{fig:scenario-4}
	\vspace{-0.75mm}
\end{figure}

\emph{Scenario I:} 
We evaluate the impact of the transmit power on problem $ \mathcal{Q} $'s objective, i.e., $ \tau $, considering $ \Gamma_{\mathrm{th},u} = \Gamma_\mathrm{th} = 4 $, $ \forall u \in \mathcal{U} $, and $ \Delta = 10^\circ $. \cref{fig:scenario-1} shows that $ \tau $ improves for all approaches as the transmit power increases, since the communication requirement remains invariant, leaving more power for sensing. In particular, {\mytextsf{OPT-SPB}} outperforms the baselines, achieving gains of over $ 27\% $, highlighting the importance of jointly designing scheduling, pairing, and beamforming, rather than addressing them separately as in the baselines. Among the baselines, {\mytextsf{BL2}} achieves the highest performance by scheduling users with the least correlated channels, effectively mitigating the high inter-user interference present in this scenario due to $ \Delta = 10^\circ $. Although {\mytextsf{BL3}} ensures optimal pairing, {\mytextsf{BL2}} outperforms it. This highlights the crucial role of effective scheduling and demonstrates that simplistic scheduling, such as the random approach used in {\mytextsf{BL3}}, can significantly degrade performance. Additionally, {\mytextsf{BL1}} underperforms compared to {\mytextsf{BL2}} , despite its focus on maximizing alignment between users and targets. This suggests that while alignment is beneficial, it is not the sole determinant of system performance and should be considered in conjunction with other factors, as demonstrated in \emph{Scenario IV}. Lastly, {\mytextsf{BL4}} performs the worst, as its scheduling and pairing are random.

\emph{Scenario II:} 
We evaluate the impact of the \gls{SINR} threshold on sensing performance, considering $ P_\mathrm{tx} = 40 $ dBm. \cref{fig:scenario-2} shows that $ \tau $ decreases for all approaches as $ \Gamma_\mathrm{th} $ increases, since the communication requirement becomes more stringent, leaving less power for sensing. {\mytextsf{OPT-SPB}} shows a significant advantage over the baselines, with gains exceeding $ 38\% $ when $ \Gamma_\mathrm{th} = 12 $. These results further emphasize the importance of joint design and highlight that meeting a stricter $ \Gamma_\mathrm{th} $ becomes more challenging with suboptimal scheduling and pairing, leading to significant sensing performance degradation.

\emph{Scenario III:} We consider the same parameters as in \emph{Scenario II} but assume $ \Delta = 30^\circ $. \cref{fig:scenario-3} shows that  {\mytextsf{OPT-SPB}} outperforms the best baseline, {\mytextsf{BL1}} in this case, by $ 33\% $. This contrasts with previous scenarios where {\mytextsf{BL2}} was the best baseline. In those scenarios, users were spaced $ \Delta = 10^\circ $ apart, leading to significant channel correlation, making user scheduling a critical factor. This favored {\mytextsf{BL2}}, whose user scheduling is designed to minimize correlation. However, in this scenario, users are spaced $ \Delta = 30^\circ $, yielding lower channel correlation, thereby reducing the importance of user scheduling and shifting the focus to pairing. This explains why {\mytextsf{BL1}} now outperforms {\mytextsf{BL2}}, as it prioritizes pairing over scheduling, using alignment maximization. Moreover, \cref{fig:scenario-2} and \cref{fig:scenario-3} show that the performance of the baselines can vary significantly depending on the scenario characteristics, with certain settings favoring some schemes over others, while {\mytextsf{OPT-SPB}} performs consistently in all scenarios.

\emph{Scenario IV:} 
\cref{fig:scenario-4} shows the beampatterns for a scenario, where $ P_\mathrm{tx} = 24 $ dBm, $ \Gamma_\mathrm{th} = 12 $, $ U = 5 $, $ T = 4 $, $ K = J = 2 $, $ \beta_1 = 50 $, $ \beta_2 = 70 $, $ \beta_3 = 90 $, $ \beta_4 = 110 $, $ \beta_5 = 130 $, $ d_1 = d_3 = 60 $ m, $ d_2 = d_4 = d_5 = 80 $ m, $ \theta_1 = 10 $, $ \theta_2 = 40 $, $ \theta_3 = 80 $, $ \theta_4 = 160 $, $ \alpha_1 = 0.04 $, $ \alpha_2 = 0.08 $, $ \alpha_3 = 0.07 $, and $ \alpha_4 = 0.04 $. Due to the high alignment between $ \mathsf{U}_1 $ and $ \mathsf{T}_2 $, they are served by a broad beam. Besides, while $ \mathsf{U}_2 $ and $ \mathsf{U}_3 $ are also in close angular proximity to $ \mathsf{T}_3 $, their longer distances from the \gls{BS} require more transmit power to serve them. Consequently, $ \mathsf{T}_3 $ is paired with $ \mathsf{U}_4 $, even though they are not aligned, because $ \mathsf{U}_4 $ is closer to the \gls{BS} and requires less power, leaving more available for sensing $ \mathsf{T}_3 $. Here, $ \mathsf{U}_4 $ is not paired with $ \mathsf{T}_1 $ or $ \mathsf{T}_4 $ because their lower reflection coefficients make them less favorable choices to maximize $ \tau $ compared to $ \mathsf{T}_3 $.
Also, three levels of phase resolution are considered: $ Q = 2 $, $ Q = 3 $, and $ Q = 4 $. Phase resolution plays a crucial role in improving the estimation of key parameters, such as \gls{AoA}, which is frequently updated in applications like tracking. Specifically, the worst \gls{CRB} of the \glspl{AoA} for the given phase resolution levels are $ 56.002 \cdot 10^{-11} $, $ 24.191 \cdot 10^{-11} $, and $ 13.480 \cdot 10^{-11} $, respectively, demonstrating that higher phase resolution leads to smaller \gls{AoA} estimation error. However, it is important to note that the \gls{CRB} serves as an ideal lower bound and may not be achievable in practice, as it relies on the existence of an estimator that satisfies specific conditions. As such, these \gls{CRB} values should be interpreted as theoretical benchmarks.

\section{Conclusions} \label{sec:conclusions}

We investigated the joint design of scheduling, pairing, and beamforming, and proposed an approach that guarantees globally optimal solutions. We compared our approach to baselines relying on heuristic scheduling and pairing, and demonstrated that such methods can vary significantly in performance, underperforming compared to our approach. Our results highlight the importance of a holistic, integrated approach that effectively addresses all design aspects simultaneously.



\appendix

\begin{proof}[\emph{\textbf{Proof of Proposition 1}}]
We use binary variables to encode the phase selection for each element of the beamforming vector. A binary vector $ \mathbf{x}_{u,n} $ is introduced for each $ \mathrm{U}_u $ and antenna element $ n $, as in $ \mathrm{D}_{1} $. Phases are selected for $ \mathrm{U}_u $ only if the user is scheduled, i.e., $ \mu_u = 1 $, which is enforced by $ \mathrm{D}_{2} $. Finally, $ \mathrm{D}_{3} $ maps $ \mathbf{x}_{u,n} $ to one of the phases in $ \mathbf{s} $.
\end{proof}

\begin{proof}[\emph{\textbf{Proof of Proposition 2}}]
	For any $ \mathrm{U}_u $, the two cases in $ \mathrm{C}_8 $ can be merged and equivalently recast as $ \widebar{\mathrm{Z}}_\mathrm{a}: \mathsf{SINR}_u \left(  \mathbf{w} \right) \geq \mu_u \cdot \Gamma_{\mathrm{th},u} $. When $ \mu_u = 1 $, then $ \widebar{\mathrm{Z}}_\mathrm{a} $ is equivalent to the first case. When $ \mu_u = 0 $, constraints $ \mathrm{D}_{2} $ and $ \mathrm{D}_{3} $ ensure $ \mathbf{w}_u = \mathbf{0} $, making $ \widebar{\mathrm{Z}}_\mathrm{a} $ collapse to the second case. Thus, constraint $ \widebar{\mathrm{Z}}_\mathrm{a} $ can be rearranged as $ \widebar{\mathrm{Z}}_\mathrm{b}: \big( \sum_{i \in \mathcal{U}, i \neq u} \big| \widetilde{\mathbf{h}}^\mathrm{H}_u \mathbf{w}_i\big|^2  + 1 \big) \mu_u \leq \Gamma_{\mathrm{th},u}^{-1} \big| \widetilde{\mathbf{h}}^\mathrm{H}_u \mathbf{w}_u \big|^2  $. Adding $ \big| \widetilde{\mathbf{h}}^\mathrm{H}_u \mathbf{w}_u \big|^2 $ to both sides of $ \widebar{\mathrm{Z}}_\mathrm{b} $ leads to $ \widebar{\mathrm{Z}}_\mathrm{c}: \big( \sum_{i \in \mathcal{U}, i \neq u} \big| \widetilde{\mathbf{h}}^\mathrm{H}_u \mathbf{w}_i\big|^2  + 1 \big) \mu_u + \big| \widetilde{\mathbf{h}}^\mathrm{H}_u \mathbf{w}_u \big|^2 \leq \big( 1 + {\Gamma_\mathrm{th},u}^{-1} \big) \big| \widetilde{\mathbf{h}}^\mathrm{H}_u \mathbf{w}_u \big|^2 $. To get rid of the couplings between variables $ \mathbf{w}_i $ and $ \mu_u $, we apply the \emph{big-M} method as in \cite[p.8]{abanto2023:radiorchestra-proactive-management-millimeter-wave-self-backhauled-small-cells-joint-optimization-beamforming-user-association-rate-selection-admission-control}, equivalently transforming $ \widebar{\mathrm{Z}}_\mathrm{c} $ into $ \widebar{\mathrm{Z}}_\mathrm{d}: \sum_{i \in \mathcal{U}} \big| \widetilde{\mathbf{h}}^\mathrm{H}_u \mathbf{w}_i\big|^2  + 1 \leq \big( 1 + \Gamma_{\mathrm{th},u}^{-1} \big) \big| \widetilde{\mathbf{h}}^\mathrm{H}_u \mathbf{w}_u \big|^2 + \big( 1 - \mu_u \big) \widebar{B}_u $, where $ \widebar{B}_u $ is an upper bound of the \gls{LHS} of $ \widebar{\mathrm{Z}}_\mathrm{d} $, computed via the trace inequality in Lemma \ref{lem:trace-inequality}. By introducing new variables $ \mathbf{W}_u \in \mathbb{C}^{N \times N} $, $ \forall u \in \mathcal{U} $, in constraint $ \mathrm{E}_{1} $, we can recast constraint $ \widebar{\mathrm{Z}}_\mathrm{d} $ as $ \mathrm{E}_{2} $, where the cyclic property of the trace, in Lemma \ref{lem:cyclic-property-trace}, was used, yielding $ \big| \widetilde{\mathbf{h}}^\mathrm{H}_u \mathbf{w}_i\big|^2 =  \mathrm{Tr} \big( \mathbf{w}_i^\mathrm{H} \widetilde{\mathbf{h}}_u \widetilde{\mathbf{h}}^\mathrm{H}_u \mathbf{w}_i \big) = \mathrm{Tr} \big(  \mathbf{W}_i \widetilde{\mathbf{H}}_u \big) $. 
\end{proof}

\begin{proof}[\emph{\textbf{Proof of Proposition 3}}]
	Given $ \mathbf{W}_u = \mathbf{w}_u \mathbf{w}^\mathrm{H}_u $, the $n$-th row of $ \mathbf{W}_u $ is $ \left[ \mathbf{w}_u \right]_n \mathbf{w}^\mathrm{H}_u $. In addition, the $ m $-th element of the $n$-th row is $\left[ \mathbf{w}_u \right]_n \left[ \mathbf{w}^{*}_u \right]_m $.
\end{proof}

\begin{proof}[\emph{\textbf{Proof of Proposition 4}}]
Replacing $ \mathrm{D}_{3} $ in $ \mathrm{F}_{1} $ results in $ \left[ \mathbf{W}_u \right]_{n,m} = \mathbf{s}^\mathrm{T} \mathbf{x}_{u,n} \mathbf{x}^\mathrm{T}_{u,m} \mathbf{s}^{*} = \mathrm{Tr} \left( \mathbf{S} \mathbf{x}_{u,n} \mathbf{x}^\mathrm{T}_{u,m} \right), \forall u \in \mathcal{U}, n, m \in \mathcal{N}. $ Since $ \mathbf{W}_u $ is Hermitian, the elements of $ \mathbf{W}_u $ are conjugate symmetrical with respect to the diagonal, whereas all the diagonal elements are $ \delta^2 $ when $ \mu_u = 1 $, and $ 0 $ when $ \mu_u = 0 $. Therefore, we only index the upper triangular part of $ \mathbf{W}_u $ resulting in $ \mathrm{G}_{1} $, $ \mathrm{G}_{2} $, and $ \mathrm{G}_{3} $. 
\end{proof}

\begin{proof}[\emph{\textbf{Proof of Proposition 5}}]
We introduce variable $ \mathbf{Y}_{u,n,m} \in \left[ 0, 1 \right]^{N \times N}  $ to replace product $ \mathbf{x}_{u,n} \mathbf{x}^T_{u,m} $, as stated in $\mathrm{H}_{1} $. Applying $ \mathrm{H}_{1} $ to $ \mathrm{G}_{1} $ leads to $ \mathrm{H}_{2} $. In addition, $ \mathbf{Y}_{u,n,m} $ is defined as having binary entries in $ \mathrm{H}_{3} $ to maintain the same nature of the product $ \mathbf{x}_{u,n} \mathbf{x}^T_{u,m} $.
\end{proof}

\begin{proof}[\emph{\textbf{Proof of Proposition 6}}]
For any scheduled user $ \mathrm{U}_u $, each of the vectors $ \mathbf{x}_{u,n} $ and $ \mathbf{x}^T_{u,m} $ have only one entry $ 1 $, according to $ \mathrm{D}_{2} $. Based on $ \mathrm{H}_{1} $, $ \mathbf{Y}_{u,n,m} $ also has only one entry $ 1 $ while the remaining entries are $ 0 $. Specifically, if $ \left[ \mathbf{x}_{u,n} \right]_l = 1 $ and $ \left[ \mathbf{x}_{u,m} \right]_i = 1 $, then $ \left[ \mathbf{Y}_{u,n,m} \right]_{l,i} = 1 $. We use this observation to eliminate the couplings $ \mathbf{x}_{u,n} \mathbf{x}^T_{u,m} $. Due to space constraints, the following explanation solely focuses on the relation between $ \mathrm{H}_{1} $ and $ \mathrm{I}_{2} $, given that $ \mathrm{I}_{1} $ and $ \mathrm{I}_{2} $ are similar in nature. Term $ \left[ \mathbf{x}_{u,n} \right]_l \mathbf{x}^T_{u,m} $ represents the $ l $-th row of $ \mathbf{Y}_{u,n,m} $, i.e., $ \left[ \mathbf{Y}_{u,n,m} \right]_{l,:} = \left[ \mathbf{x}_{u,n} \right]_l \mathbf{x}^T_{u,m} $. When $ \left[ \mathbf{x}_{u,n} \right]_l = 0 $, all the elements of $ \left[ \mathbf{Y}_{u,n,m} \right]_{l,:} $ are $ 0 $, meaning that the sum of all the elements of $ \left[ \mathbf{Y}_{u,n,m} \right]_{l,:} $ is $ \left[ \mathbf{x}_{u,n} \right]_l = 0 $. When $ \left[ \mathbf{x}_{u,n} \right]_l = 1 $, then $ \left[ \mathbf{Y}_{u,n,m} \right]_{l,:} = \mathbf{x}^T_{u,m} $. Given that $ \mathbf{x}^T_{u,m} $ has only one entry equal to $ 1 $, then the sum of elements of $ \left[ \mathbf{Y}_{u,n,m} \right]_{l,:} $ must be $ \left[ \mathbf{x}_{u,n} \right]_l = 1 $. Since $ \mathbf{Y}_{u,n,m} $ has one entry $ 1 $, we can relax $ \mathrm{H}_3 $ such that $ \left[ \mathbf{Y}_{u,n,m} \right]_{l,i} $ is continuous in $ \left[ 0, 1 \right] $, which does not alter the solution space of $ \mathbf{Y}_{u,n,m} $. In particular, this is ensured by the total unimodularity principle \cite[ch.13]{papadimitriou1998:combinatorial-optimization-algorithms-complexity}, which allows recasting $ \mathrm{H}_3 $ and $ \mathrm{I}_3 $, as long as $ \mathrm{I}_1 $ and $ \mathrm{I}_2 $ are included.
\end{proof}

\begin{proof}[\emph{\textbf{Proof of Proposition 7}}]

By merging the two cases in $ \mathrm{C}_{10} $ and leveraging the cyclic property of the trace, in Lemma \ref{lem:cyclic-property-trace}, $ \mathrm{C}_{10} $ can be expressed as constraint $ \widebar{\mathrm{Z}}_\mathrm{e}: \mathrm{Tr} \big( \mathbf{G}_t \mathbf{v}_t \mathbf{v}^\mathrm{H}_t \big) \geq \tau \lambda_t $ for any target $ \mathrm{T}_t $. While the first case in $ \mathrm{C}_{10} $ is directly implied by $ \widebar{\mathrm{Z}}_\mathrm{e} $, the second case can be derived from $ \mathrm{C}_{6} $ and $ \mathrm{C}_{13} $, resulting in the equality to $ 0 $. Replacing $ \mathbf{v}_t $, in $ \mathrm{C}_{13} $, into $ \widebar{\mathrm{Z}}_\mathrm{e} $ leads to $ \widebar{\mathrm{Z}}_\mathrm{f}: \mathrm{Tr} \big( \mathbf{G}_t \textstyle \big( \sum_{u \in \mathcal{U}} \mathbf{w}_u \rho_{u,t} \big) \big( \sum_{i \in \mathcal{U}} \mathbf{w}_i \rho_{i,t} \big)^\mathrm{H} \big) \geq \tau \lambda_t $, eliminating the need for $ \mathbf{v}_t $. Based on $ \mathrm{C}_{6} $ and $ \mathrm{C}_{7} $, product $ \rho_{u,t} \rho_{i,t} $ is $ 0 $ when $ u \neq i $ because a target can only be paired with one user, thus leading us to defining $ \widebar{\mathrm{Z}}_\mathrm{f} $ as $ \widebar{\mathrm{Z}}_\mathrm{g}: \textstyle \sum_{u \in \mathcal{U}} \mathrm{Tr} \big( \mathbf{G}_t \mathbf{W}_u \rho_{u,t} \big) \geq \tau \lambda_t $, since $ \rho_{u,t}^2 = \rho_{u,t} $. Constraint $ \widebar{\mathrm{Z}}_\mathrm{g} $ exhibits two couplings, specifically, $ \mathbf{W}_u $ with $ \rho_{u,t} $ and $ \tau $ with $  \lambda_t $, complicating the optimization of these variables. Leveraging $ \mathrm{C}_{5} $ and applying the \emph{big-M} method to $ \widebar{\mathrm{Z}}_\mathrm{g} $, as in \cite[p.8]{abanto2023:radiorchestra-proactive-management-millimeter-wave-self-backhauled-small-cells-joint-optimization-beamforming-user-association-rate-selection-admission-control}, we get rid of these multiplicative couplings, and thus express $ \widebar{\mathrm{Z}}_\mathrm{g} $ equivalently as $ \mathrm{J}_\mathrm{1} $, where $ \widebar{D}_t $ is an upper bound of the \gls{LHS} of $ \widebar{\mathrm{Z}}_\mathrm{g} $, computed via Lemma~\ref{lem:trace-inequality}. 
\end{proof}

\begin{proof}[\emph{\textbf{Proof of Proposition 8}}]

The procedure follows a similar approach to Proposition 7. Note that the first case of $ \mathrm{C}_{11} $ is only active when $ \lambda_t = \lambda_q = 1 $, while the second case is always satisfied due to the limited transmit power. Therefore, instead of using $ \infty $ in the second case, we can substitute it with the upper bound $ \widebar{D}_t $, derived in Proposition 7. Thus, for any target $ \mathrm{T}_t $, this allows us to combine the two cases of $ \mathrm{C}_{11} $ into a single case, resulting in  $ \widebar{\mathrm{Z}}_\mathrm{h}: \textstyle \mathbf{v}^\mathrm{H}_t \mathbf{G}_q \mathbf{v}_t \leq \xi_\mathrm{th} + \left(1 - \lambda_t \lambda_q \right) \widebar{D}_t $. The \gls{LHS} of $ \widebar{\mathrm{Z}}_\mathrm{h} $ is similar to the \gls{LHS} of $ \widebar{\mathrm{Z}}_\mathrm{e} $, thus we can recast $ \widebar{\mathrm{Z}}_\mathrm{h} $ as $ \widebar{\mathrm{Z}}_\mathrm{i}: \textstyle \sum_{u \in \mathcal{U}} \mathrm{Tr} \big( \mathbf{G}_q \mathbf{W}_u \rho_{u,t} \big) \leq \xi_\mathrm{th} + \left(1 - \lambda_t \lambda_q \right) \widebar{D}_t $. This constraint can be split into $ \widebar{\mathrm{Z}}_{\mathrm{i}_1}: \pi_{t,q} = \lambda_t \lambda_q $ and $ \widebar{\mathrm{Z}}_{\mathrm{i}_2}: \textstyle \sum_{u \in \mathcal{U}} \mathrm{Tr} \big( \mathbf{G}_q \mathbf{W}_u \rho_{u,t} \big) \leq \xi_\mathrm{th} + \left(1 - \pi_{t,q} \right) \widebar{D}_t $. Leveraging Lemma \ref{lem:product-binary-variables}, $ \widebar{\mathrm{Z}}_{\mathrm{i}_1} $ can be expressed as $ \mathrm{K}_1 $, $ \mathrm{K}_2 $, $ \mathrm{K}_3 $, and $ \mathrm{K}_4 $, while $ \widebar{\mathrm{Z}}_{\mathrm{i}_2} $ is equivalently recast as $ \mathrm{K}_5 $ using the \emph{big-M} method as in Proposition 7.
\end{proof}

\begin{lemma} \label{lem:trace-inequality}
	\cite[p.2]{yang2002:note-trace-inequality-products-hermitian-matrix-power} Given positive semidefinite matrices $ \mathbf{A} $ and $ \mathbf{B} $, the following holds $ \mathrm{Tr} \left( \mathbf{A} \mathbf{B} \right) \leq \mathrm{Tr} \left( \mathbf{A} \right) \mathrm{Tr} \left( \mathbf{B} \right)$. 
\end{lemma}

\begin{lemma} \label{lem:cyclic-property-trace}
	Given matrices $ \mathbf{A} $, $ \mathbf{B} $, $ \mathbf{C} $, and $ \mathbf{D} $ of compatible dimensions, the following holds $ \mathrm{Tr} \left( \mathbf{A} \mathbf{B} \mathbf{C} \mathbf{D} \right) = \mathrm{Tr} \left( \mathbf{B} \mathbf{C} \mathbf{D} \mathbf{A}  \right)$. 
\end{lemma}

\begin{lemma} \label{lem:product-binary-variables}
	Let $ c = a b $, where $ a, b \in \left\lbrace 0, 1 \right\rbrace $. Thus, this equality is equivalent to the intersection of $ \mathrm{R}_1: c \leq a $, $ \mathrm{R}_2: c \leq b $, $ \mathrm{R}_3: c \geq a + b -1 $, and $ \mathrm{R}_4: c \in \left[ 0, 1 \right] $.
\end{lemma}

\bibliographystyle{IEEEtran}
\bibliography{IEEEabrv,ref}


\end{document}